\documentclass[11pt,a4paper]{article}
\pdfoutput=1
\usepackage{jcappub}

\usepackage{appendix}
\usepackage{array}
\newcolumntype{x}[1]{>{\centering\arraybackslash}p{#1}}

\usepackage{epsfig}
\usepackage{graphicx}
\usepackage{dcolumn}
\usepackage{amsmath}
\usepackage{enumerate}
\def\lsim{\mathrel{\hbox{\rlap{\hbox{\lower4pt\hbox{$\sim$}}}\hbox{$<$}}}}

\makeatletter

\newcommand{\Rmnum}[1]{\expandafter\@slowromancap\romannumeral #1@}
\makeatother

\title{Low reheating temperatures in monomial and binomial inflationary models}

\author[a]{Thomas Rehagen}
\author[a]{and Graciela B. Gelmini}

\affiliation[a]{Department of Physics and Astronomy, UCLA,\\
475 Portola Plaza, Los Angeles, CA 90095, USA}

\emailAdd{trehagen@physics.ucla.edu}
\emailAdd{gelmini@physics.ucla.edu}


\abstract{We investigate the allowed range of reheating temperature values in light of the Planck 2015 results and the recent joint analysis of Cosmic Microwave Background (CMB) data from the BICEP2/Keck Array and Planck experiments, using monomial and binomial inflationary potentials.  While the well studied $\phi^2$ inflationary potential is no longer favored by current CMB data, as well as $\phi^p$ with $p>2$, a $\phi^1$ potential and canonical reheating ($w_{re}=0$) provide a good fit to the CMB measurements.  In this last case, we find that the Planck 2015 $68\%$ confidence limit upper bound on the spectral index, $n_s$, implies an upper bound on the reheating temperature of $T_{re}\lesssim 6\times 10^{10}\,{\rm GeV}$, and excludes instantaneous reheating.  The low reheating temperatures allowed by this model open the possiblity that dark matter could be produced during the reheating period instead of when the Universe is radiation dominated, which could lead to very different predictions for the relic density and momentum distribution of WIMPs, sterile neutrinos, and axions.  We also study binomial inflationary potentials and show the effects of a small departure from a $\phi^1$ potential.  We find that as a subdominant $\phi^2$ term in the potential increases, first instantaneous reheating becomes allowed, and then the lowest possible reheating temperature of $T_{re}=4\,{\rm MeV}$ is excluded by the Planck 2015 $68\%$ confidence limit.
}

\begin{document}

\maketitle

\section{Introduction}

Dark matter candidates, such as Weakly Interacting Massive Particles (WIMPs), sterile neutrinos, and axions are produced before Big Bang Nucleosynthesis (BBN).  In order to preserve the thermal history of the Universe, including BBN and all subsequent events, the lower limit imposed by BBN on the highest temperature of the radiation dominated epoch (in which BBN happens), i.e. the reheating temperature, is $4\,{\rm MeV}$ \cite{Hannestad:2004px} (see also \cite{DeBernardis:2008zz}).  Notice that, for example, the standard freeze-out temperature of thermal WIMPs is $T_{fo}\simeq m/20$, thus WIMPs with mass $m>80\,{\rm MeV}$ are produced at temperatures above the BBN limit.  

We do not know the history of the Universe at temperatures above the BBN lower limit.  Thus, in order to compute the relic abundance of dark matter candidates, assumptions must be made about the history of the Universe prior to BBN.  The standard assumption is that the Universe was radiation dominated up to very large temperatures.  However, there are other non-standard pre-BBN cosmological models, such as low reheating temperature models, that make different assumptions about the history of the Universe.  These models make predictions of the relic density (see e.g. Ref.~\cite{Gelmini:2006pw}) and momentum distribution (see e.g. Ref.~\cite{Gelmini:2006vn}) of dark matter candidates that can be very different from the predictions made under the assumption that the dark matter is produced when the Universe is dominated by radiation.  

Recent experiments studying the Cosmic Microwave Background (CMB), such as the Background Imaging of Cosmic Extragalactic Polarization 2 (BICEP2) and Planck satellite experiments, are searching for evidence of primordial gravitational waves generated during inflation.  While the BICEP2 experiment claimed the potential detection of a high level of primordial tensor modes in the polarization pattern of the CMB~\cite{Ade:2014xna}, the expected level of tensor modes shifted downward as the contribution due to foreground dust was better constrained.  Indeed, a recent joint analysis of BICEP2/Keck Array \cite{Ade:2015fwj} and Planck data has put an upper limit on the tensor-to-scalar ratio of amplitudes of perturbations produced during inflation at the scale (so called pivot scale) $k=0.05\,{\rm Mpc^{-1}}$, \cite{Ade:2015tva}
\begin{equation}
r<0.12
\label{rval}
\end{equation}
at the $95\%$ confidence level.  This limit can be combined with the already measured amplitude of the scalar perturbations to find a bound on the energy scale of the inflationary potential $V$ of $V^{1/4}<1.2\times 10^{16}\,{\rm GeV}$.

Inflation is a period of rapid expansion of the Universe before it became dominated by radiation.  The usual inflationary models consist of a scalar field slowly rolling down an almost flat potential (see e.g. \cite{Guth:1980zm,Linde:1983gd}).  Reheating happens when the slow-roll period ends and the energy of the inflaton field is converted into radiation (see \cite{Allahverdi:2010xz} for a review).  If reheating is assumed to be instantaneous, then the reheating temperature, $T_{re}$, is given by
\begin{equation}
V\simeq \frac{\pi^2}{30}g_{\star}(T_{re})T_{re}^4.
\label{inst}
\end{equation}
Assuming that the degrees of freedom come from the particles in the Standard Model ($g_\star\simeq 107$ for $T\gtrsim 175\,{\rm GeV}$), Eq.~\ref{inst} gives $T_{re}\lesssim 7\times 10^{15}\,{\rm GeV}$.  If reheating is instantaneous, the Universe would be dominated by radiation up to this very large temperature.

However, the physics of reheating is highly uncertain, and the reheating period may be extended.  During this extended period, the energy density of the Universe decreases due to the expansion of the Universe, and is finally converted to a radiation bath at a smaller temperature than that given by Eq.~\ref{inst} (see e.g.~\cite{Martin:2006rs,Martin:2010kz, Martin:2014vha, Dai:2014jja,Drewes:2013iaa}).  This opens up the possibility that the reheating temperature may not be much larger than the limit of $4\,{\rm MeV}$ imposed by BBN.

Here we explore the possible range of reheating temperature values implied by some simple inflationary and reheating models, in light of the Planck 2015 results and the recent BICEP2/Keck Array and Planck joint analysis.  In particular, we want to test whether $T_{re}$ can approach the BBN limit in any of these models.  We first consider the archetypal models of inflation, large field inflation models, which are characterized by a monomial potential.  The simplest model, $V=m^2\phi^2$ \cite{Belinsky:1985zd}, has received ample attention recently because it fitted the previous data well \cite{Creminelli:2014oaa,Dai:2014jja}.  Thus, we consider similar monomial ($V\propto \phi^p$) inflationary potentials and binomial ($V\propto \phi^p+b\phi^q$) inflationary potentials, where one of the two terms is dominant.  We consider inflationary models with potentials of $m^2 \phi^2$, $\lambda \phi^1$ (motivated by axion monodromy models \cite{McAllister:2008hb}), and $\lambda \phi^4$ (as in chaotic inflation \cite{Linde:1983gd}), and for the first time, as a modification of the monomial potential, $\phi^p+b\phi^q$, for general $p$ and $q$, but $b\phi^{q-p} \ll1$.

The reheating period can be described with an effective equation of state for a cosmic fluid $w_{re}=P/\rho$ (see e.g. Ref.~\cite{Martin:2006rs}), where $P$ is the pressure of the fluid, $\rho$ is the energy density, and $w_{re}$ is the equation of state parameter.  The conservation of energy equation, ${\rm d}(\rho a^3)=-P{\rm d}a^3,$ where $a$ is the scale factor, then implies that the energy density scales as $\rho \propto a^{-3(1+w_{re})}$.  For a Universe dominated by the oscillations of a scalar field around the minimum of a monomial potential, $\phi^p$, during reheating (which can be different than the potential during inflation), $w_{re}=(p-2)/(p+2)$ \cite{Turner:1983he}.   In the canonical model of reheating \cite{Abbott:1982hn,Dolgov:1982th,Albrecht:1982mp}, the inflaton field oscillates about the minimum of a quadratic potential, and decays to relativistic particles.  In this case, $w_{re}=0$. As in Refs.~\cite{Martin:2006rs,Martin:2010kz,Martin:2014vha, Dai:2014jja}, we use a range of $w_{re}$ values to parametrize the physics of the reheating period.  Inflation ends when $w_{re}\simeq -1/3$, so during reheating, the equation of state parameter must be greater than $w_{re}\gtrsim -1/3$.   In addition, for scalar fields $\rho\geq |P|$, thus $w_{re}\leq 1$.  In this paper, we consider models with $w_{re}=0$, $w_{re}=1/3$ (corresponding to $V\propto \phi^4$ during reheating), and an intermediary value, $w_{re}=1/6$, as an approximate sampling of the reasonable range of $w_{re}$.  For example, Ref.~\cite{Podolsky:2005bw} find values of $w_{re}$ between $0$ and $0.25,$ including preheating.   We also consider two extreme exotic models with $w_{re}=-1/3$ (as at the end of inflation) and $w_{re}=2/3$ (corresponding to $V\propto \phi^{10}$).

The characteristics of inflation are already constrained by measurements of CMB anisotropies and large scale structures originating from quantum fluctuations of the inflaton and gravitational fields during inflation(see e.g. \cite{Mukhanov:1981xt}).   In addition to the BICEP2/Keck Array and Planck joint analysis bound on $r$, given in Eq.~\ref{rval}, we use the Planck 2015 measurement \cite{Planck:2015xua} of the spectral index
\begin{equation}
n_s=0.9655\pm 0.0062,
\label{nsval}
\end{equation}
and the primordial scalar amplitude
\begin{equation}
{\rm ln}(10^{10}A_s)=(3.089\pm 0.036)
\label{Asval},
\end{equation}
to constrain the shape of the potential.  The measurements of $n_s$ and $A_s$ are given here with $68\%$ confidence limits.  During the slow roll phase, $n_s$, $A_s$, and $r$ can be related to the height of the inflationary potential, $V$, and its first and second derivatives ($V_\phi$ and $V_{\phi \phi}$) by
\begin{equation}
n_s-1\simeq \frac{M_P^2}{V^2}\left(2V_{\phi \phi} V-3V_{\phi}^2\right),
\end{equation}
\begin{equation}
A_s\simeq \frac{V^3}{12\pi^2 M_P^6 V_{\phi}^2},
\end{equation}
and
\begin{equation}
r\simeq 8 M_P^2\frac{V_{\phi}^2}{V^2},
\end{equation}
where $M_P=2.4\times 10^{18}\,{\rm GeV}$ is the reduced Planck mass.  During the slow roll phase, the parameters $\epsilon_V$ and $\eta_V$, given by
\begin{equation}
\epsilon_V\equiv \frac{M_P^2}{2}\left( \frac{V_\phi}{V}\right)^2
\end{equation}
and
\begin{equation}
\eta_V\equiv M_P^2\frac{V_{\phi \phi}}{V}
\end{equation}
are much smaller than 1.

We will compute $n_s$ and $r$ in terms of the reheating temperature for a given inflationary potential and reheating model (parametrized by $w_{re}$).  In order to do so, we start by relating $T_{re}$ with $N_k$ and $\rho_{end}$, where $N_k$ is the number of e-folds of expansion of the scale factor of the Universe between the moment when the pivot scale, $k$, exits the horizon and the end of inflation, and $\rho_{end}$ is the energy density at the end of inflation.  This relation depends on the reheating model (namely $w_{re}$).  Assuming that the energy density does not change appreciably during inflation, $\rho_{end}$ is given by the product $rA_s$.  We then compute $N_k$ in terms of $n_s$ and $r$ for each particular inflationary potential.  Our work extends the analyses carried out in Refs.~\cite{Martin:2006rs,Martin:2010kz, Martin:2014vha, Creminelli:2014oaa, Dai:2014jja, Munoz:2014eqa} by including, for the first time, an analytical treatment of binomial inflationary potentials, as well as monomial potentials, by taking a larger range of values of $w_{re}$, and by including the BICEP2/Keck Array and Planck joint analysis bound on $r$ as a constraint.  Our approach is complementary to that of Ref.~\cite{Domcke:2015iaa}, which studies the relationship between the inflaton decay rate and $T_{re}$.

We find particularly interesting that, while the well studied $\phi^2$ inflationary potential reproduces the value of $n_s$ favored by Planck, the upper limit on $r$ found by the BICEP2/Keck Array and Planck joint analysis cannot be satisfied with this potential by any of the reheating models we consider, in agreement with Fig.~21 of Ref.\cite{Planck:2015xua}.  The quadratic potential, therefore, is no longer favored; a $\phi^1$ potential, however, does provide a good fit to the CMB measurements.  For this potential, the Planck 2015 $68\%$ confidence limit on $n_s$ and the upper bound on $r$ imply an upper bound $T_{re}\lesssim 6\times 10^{10}\,{\rm GeV}$ for the reheating temperature, assuming canonical reheating.  Only exotic reheating models with $w_{re}<0$, such as $w_{re}=-1/3$, allow for higher reheating temperatures, with a $\phi^1$ inflationary potential.  If the Universe has a low reheating temperature, as allowed by this model, the possibility is opened that dark matter could be produced during the reheating period instead of when the Universe is radiation dominated, which could lead to very different predictions for the relic density and even momentum distribution of WIMPs, sterile neutrinos, and axions from the predictions assuming that dark matter is produced when the Universe is radiation dominated.

\section{Relationship between $N_k$ and $T_{re}$}

In this section, we rederive a relation between the number of e-folds, $N_k={\rm ln}\,(a_{end}/a_k),$ where $a_{end}$ and $a_k$ are the scale factors at the end of inflation and when the pivot scale exited the horizon, respectively, and the reheating temperature, $T_{re}$ in terms of known quantities $k$, $r$, $A_s$, $a_{eq}$, and $\rho_{eq}$ (the last two are the scale factor and energy density at matter-radiation equality, when the energy densities of matter and radiation are equal).  We do this by tracing the dilution of the energy density of the Universe during the reheating phase.

During reheating, the energy density, $\rho$, scales as $\rho\propto a^{-3(1+w_{re})}$, where $w_{re}$ is the equation of state parameter.   This means that
\begin{equation}
\frac{\rho_{end}}{\rho_{re}}=\left(\frac{a_{re}}{a_{end}} \right)^{3(1+w_{re})},
\label{rhore}
\end{equation}
where $\rho_{end}$, $\rho_{re}$ and $a_{end}$, $a_{re}$ are the energy density and scale factor at the end of inflation and at the end of reheating, respectively.  After reheating comes an epoch of radiation domination (with $\rho\propto a^{-4}$), which lasts until the Universe becomes matter dominated at matter-radiation equality.  Then
\begin{equation}
\frac{\rho_{re}}{\rho_{eq}}=\left(\frac{a_{eq}}{a_{re}} \right)^{4}.
\label{rhorad}
\end{equation}
Combining Eqs.~\ref{rhore} and \ref{rhorad}, we find
\begin{equation}
\frac{a_{end}}{a_{eq}}=\left(\frac{\rho_{eq}}{\rho_{end}} \right)^{\frac{1}{3(1+w_{re})}}\left(\frac{\rho_{eq}}{\rho_{re}} \right)^{\frac{3w_{re}-1}{12(w_{re}+1)}}.
\label{comb}
\end{equation}
Since $a_{end}=a_k\, {\rm exp}(N_k),$ we can replace $a_{end}$ in Eq.~\ref{comb} and solve for $N_k$,
\begin{equation}
N_k={\rm ln}\left(\frac{a_{eq}}{a_{k}} \right)+\frac{1}{3(1+w_{re})}\,{\rm ln}\left(\frac{\rho_{eq}}{\rho_{end}} \right)+\frac{3w_{re}-1}{12(w_{re}+1)}\,{\rm ln}\left(\frac{\rho_{eq}}{\rho_{re}} \right).
\label{Nk}
\end{equation}
This expression is valid for any inflationary potential.  This potential is needed to relate $\rho_{end}$ to the energy density at the pivot scale $\rho_{k}$.  An equation similar to Eq.~\ref{Nk} is given originally in Eq.~15 of Ref.~\cite{Martin:2010kz}, and again (but solving for $N_{re}={\rm ln}(a_{re}/a_{end})$ and applied only to the case of a monomial inflationary potential) in Eq.~11 of Ref.~\cite{Dai:2014jja}.

The energy density at matter-radiation equality is $\rho_{eq}\simeq 2 \rho_{rad}=2\left((\pi^2/30)g_{\star,eq} T^4_{eq}\right),$ where $T_{eq}=(a_0/a_{eq})T_0$ and $g_{\star,eq}=3.36$ are the temperature and number of relativistic degrees of freedom at matter-radiation equality, respectively.  The energy density at the end of the reheating epoch is related to the reheating temperature by $\rho_{re}=(\pi^2/30)g_{\star,re}T_{re}^4$.  The number of relativistic degrees of freedom at the end of reheating, $g_{\star,re}$ depends on the reheating temperature, $T_{re},$ and is approximately given by $g_{\star,re}=107,$  $90,$ and $11$ for $T_{re}\gtrsim175\,{\rm GeV}$, $175\,{\rm GeV}\gtrsim T_{re} \gtrsim 200\,{\rm MeV}$, and $200\,{\rm MeV}\gtrsim T_{re} \gtrsim 1\,{\rm MeV},$ respectively, where $T\simeq 200\,{\rm MeV}$ roughly corresponds to the temperature of the QCD phase transition, and $T\lesssim 175\,{\rm GeV}$  is when the temperature drops below the top quark mass \cite{Beringer}.

There are still two unknown parameters, $a_k$ and $\rho_{end}$, in  Eq.~\ref{Nk}.  $a_k$ can be related to $k$, $r$, and $A_s$ because at the moment the comoving scale $k$ exits the horizon, the Hubble expansion rate, $H_k\simeq \sqrt{(\pi^2/2)M_P^2rA_s},$ is related to $k$ and $a_k$ by $H_k=k/a_k$.  Although we do not know $\rho_{end}$, we do know the energy density when the pivot scale exits the horizon, $\rho_k=3M_P^2H_k^2\simeq (3\pi^2/2)M_P^4rA_s$.  Given a particular inflationary potential, we can calculate the relationship between $r$ and $n_s$, which will allow us to find the values of $H_k$ and $\rho_{k}$ by using the Planck measurement of $n_s$ given in Eq.~\ref{nsval}.  If the energy density during inflation is approximately constant, then $\rho_{end}\simeq \rho_k$.  As we will show later, this approximation introduces an error of no more than $6\%$ to the value of $N_k$ in the models with $w_{re}=0$ that we consider in subsequent sections.

\begin{figure}
\begin{center}
\includegraphics[width=10cm]{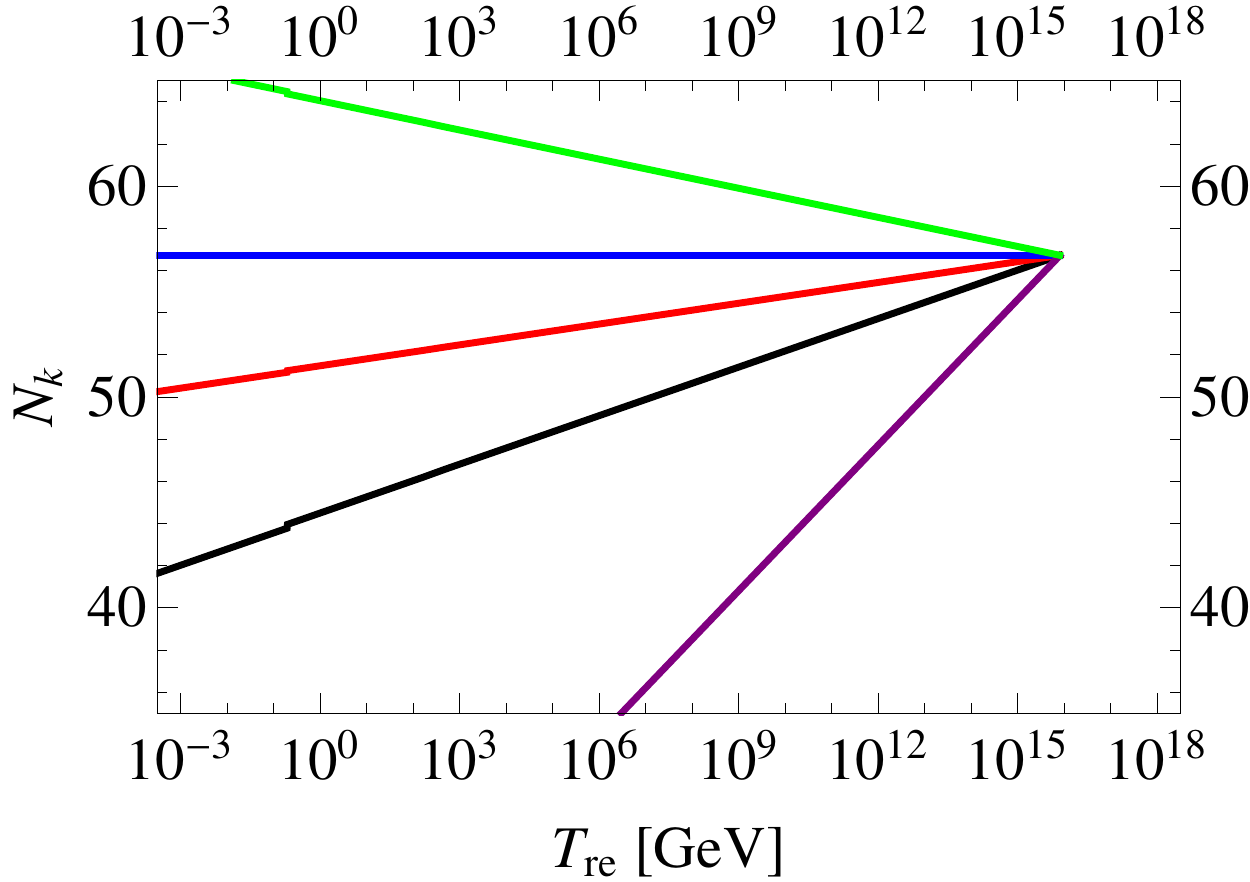}
\caption{$N_k$ for pivot scale $k=0.05\,{\rm Mpc}^{-1}$ as a function of the reheating temperature $T_{re}$ for $w_{re}=0$ (black), $w_{re}=1/6$ (red), $w_{re}=1/3$ (blue), $w_{re}=-1/3$ (purple), and $w_{re}=2/3$ (green), using $n_s=0.9655$, $A_s=2.2\times10^{-9},$ and $a_{eq}/a_0=3360.$  To make this figure, we assumed that the inflationary potential is quadratic.}
\label{NTPlot}
\end{center}
\end{figure}

Fig.~\ref{NTPlot} shows $N_k$ as given in Eq.~\ref{Nk} as a function of the reheating temperature $T_{re}$ for $w_{re}=0$ (black), $w_{re}=1/6$ (red), $w_{re}=1/3$ (blue), $w_{re}=-1/3$ (purple), and $w_{re}=2/3$ (green), for $k=0.05\,{\rm Mpc}^{-1}$, the central values of the measurements of $n_s$ and $A_s$, given in Eqs.~\ref{nsval} and \ref{Asval}, and $a_{eq}/a_0=3360$ \cite{Beringer}.  To make this figure, we assumed that the inflationary potential is quadratic.  We note that taking values of $n_s$ and $A_s$ at the edge of the $68\%$ confidence limit ranges given in Eqs.~\ref{nsval} and \ref{Asval}, respectively, will only change the calculated value of $N_k$ by $\Delta N_k\simeq 0.1$.  The plot ends at $T_{re}\simeq 7\times 10^{15}\,{\rm GeV}$ because this reheating temperature occurs when the inflaton field decays instantaneously into radiation, thus $\rho_{end}$ does not redshift before being transformed into radiation.  The case of $w_{re}=1/3$ gives a straight line because the expansion rate in the reheating epoch is the same as in a radiation dominated epoch.  In this case, it does not matter when the Universe becomes radiation dominated, because it expands in the same way, i.e. $\rho \propto a^{-4},$ both before and after the transition.

Fig.~\ref{NTPlot} makes it clear that if the reheating temperature is known, the number of e-folds from when the pivot point exits the horizon to the end of inflation can be calculated, given an inflationary potential and a reheating model.  As previously mentioned, the lowest possible reheating temperature is $T_{re}\simeq 4\,{\rm MeV}$.  As can be seen in Fig.~\ref{NTPlot}, this means that $N_k\simeq 42.5$ is the lowest allowed number of e-folds between the time when the pivot scale of $k=0.05\,{\rm Mpc}^{-1}$ exits the horizon and the end of inflation for $w_{re}=0$, assuming a quadratic inflationary potential.  Lower bounds on $N_k$ for other values of $w_{re}$ can be found similarly.  The upper bound for all reasonable reheating models with $w_{re}\leq 1/3$ is $N_k\lesssim 57.$

We made Fig.~\ref{NTPlot} under the assumption that the energy density is approximately constant during inflation.  If the energy density does change during inflation, i.e. $\rho_k>\rho_{end}$, we must add the term
\begin{equation}
\Delta N_k = \frac{1}{3(1+w_{re})}{\rm ln}\left( \frac{\rho_{k}}{\rho_{end}} \right)
\end{equation}
 to the right side of Eq.~\ref{Nk}.  In the models we consider in the subsequent sections, the ratio of the energy density when the pivot scale leaves the horizon to the energy density at the end of inflation varies from $\rho_k/\rho_{end}\simeq 15$ to $\rho_k/\rho_{end}\simeq 3700$, which leads to a shift of $\Delta N_k\simeq 0.9$ to $\Delta N_k \simeq 2.7$ for $w_{re}=0$.  A similar estimation can be made for other values of $w_{re}$, and $\Delta N_k$ is always small.

\section{Constraints on $T_{re}$ from $n_s$ and $r$}

In this section we find expressions for the spectral index, $n_s$, and the tensor-to-scalar ratio, $r$, for monomial ($V\propto \phi^p$) and binomial ($V=\phi^p+b\phi^q$) potentials, as a function of $N_k.$  We then use the ranges of $r$ and $n_s$, given in Eqs.~\ref{rval} and \ref{nsval} respectively, to derive constraints on $N_k,$ and subsequently $T_{re}.$

\subsection{Monomial potentials}

The simplest models of the inflaton potential are monomial potentials.  These potentials have the form
\begin{equation}
V(\phi)=\lambda M_P^4 \left(\frac{\phi}{M_P} \right)^p.
\end{equation}
The slow roll parameters $\epsilon_V$ and $\eta_V$ are given by
\begin{equation}
\epsilon_V\equiv \frac{M_P^2}{2}\left( \frac{V_\phi}{V}\right)^2=\frac{p^2}{2}\left(\frac{M_p}{\phi}\right)^2,
\end{equation}
and
\begin{equation}
\eta_V\equiv M_P^2\frac{V_{\phi \phi}}{V}=p(p-1)\left( \frac{M_P}{\phi}\right)^2.
\end{equation}
At the end of inflation, i.e. when the slow roll approximations break down, the first slow roll parameter is generally taken to be $\epsilon_V=\epsilon_{end}\simeq 1$.  Thus the value of the inflaton field at the end of inflation is
\begin{equation}
\frac{\phi_{end}}{M_P}=\frac{p}{\sqrt{2\epsilon_{end}}}.
\end{equation}

When the slow roll approximation holds, 
\begin{equation}
N_k \simeq \frac{1}{M_P^2}\int^{\phi_k}_{\phi_{end}} \frac{V}{V_\phi}d\phi,
\label{Nkdef}
\end{equation}
where $M_P$ is the reduced Planck mass, $V$ is the inflaton potential, $V_{\phi}$ is the derivative of the potential, and $\phi_k$ and $\phi_{end}$ are the values of the inflaton field when the pivot scale exits the horizon and at the end of inflation, respectively.

Using Eq.~\ref{Nkdef}, we can then find an expression for $N_k$ in the case of monomial potentials
\begin{equation}
N_k=\frac{1}{2p}\left[ \left(\frac{\phi_k}{M_P}\right)^2-\frac{p^2}{2\epsilon_{end}}\right].
\end{equation}
This relation can be inverted to solve for the value of the inflaton field when the pivot scale exits the horizon
\begin{equation}
\left(\frac{\phi_k}{M_P}\right)^2=2pN_k+\frac{p^2}{2\epsilon_{end}},
\end{equation}
and finally we find expressions for $\epsilon_V$ and $\eta_V$ as a function of $N_k$
\begin{equation}
\epsilon_V=\frac{p\epsilon_{end}}{4N_k\epsilon_{end}+p}
\label{epsformm}
\end{equation}
and
\begin{equation}
\eta_V=\frac{2(p-1)\epsilon_{end}}{4N_k\epsilon_{end}+p}.
\label{etaformm}
\end{equation}

The spectral index, $n_s,$ and the tensor-to-scalar ratio, $r,$ are given in terms of the slow roll parameters,
\begin{equation}
n_s-1 \simeq 2\eta_V-6\epsilon_V,
\label{nsm}
\end{equation}
and
\begin{equation}
r\simeq 16\epsilon_V,
\label{rm}
\end{equation}
which for monomial potentials gives
\begin{equation}
n_s-1\simeq \frac{-2(p+2)\epsilon_{end}}{4N_k\epsilon_{end}+p},
\label{nsma}
\end{equation}
and
\begin{equation}
r\simeq \frac{16p\epsilon_{end}}{4N_k \epsilon_{end}+p}.
\label{rma}
\end{equation}
We can see from Eqs.~\ref{nsma} and \ref{rma} that for monomial inflationary potentials, the tensor-to-scalar can be related to the spectral index by
\begin{equation}
n_s-1=\frac{-(p+2)r}{8p}.
\end{equation}

 \begin{figure} 
\begin{center}
\includegraphics[width=10cm]{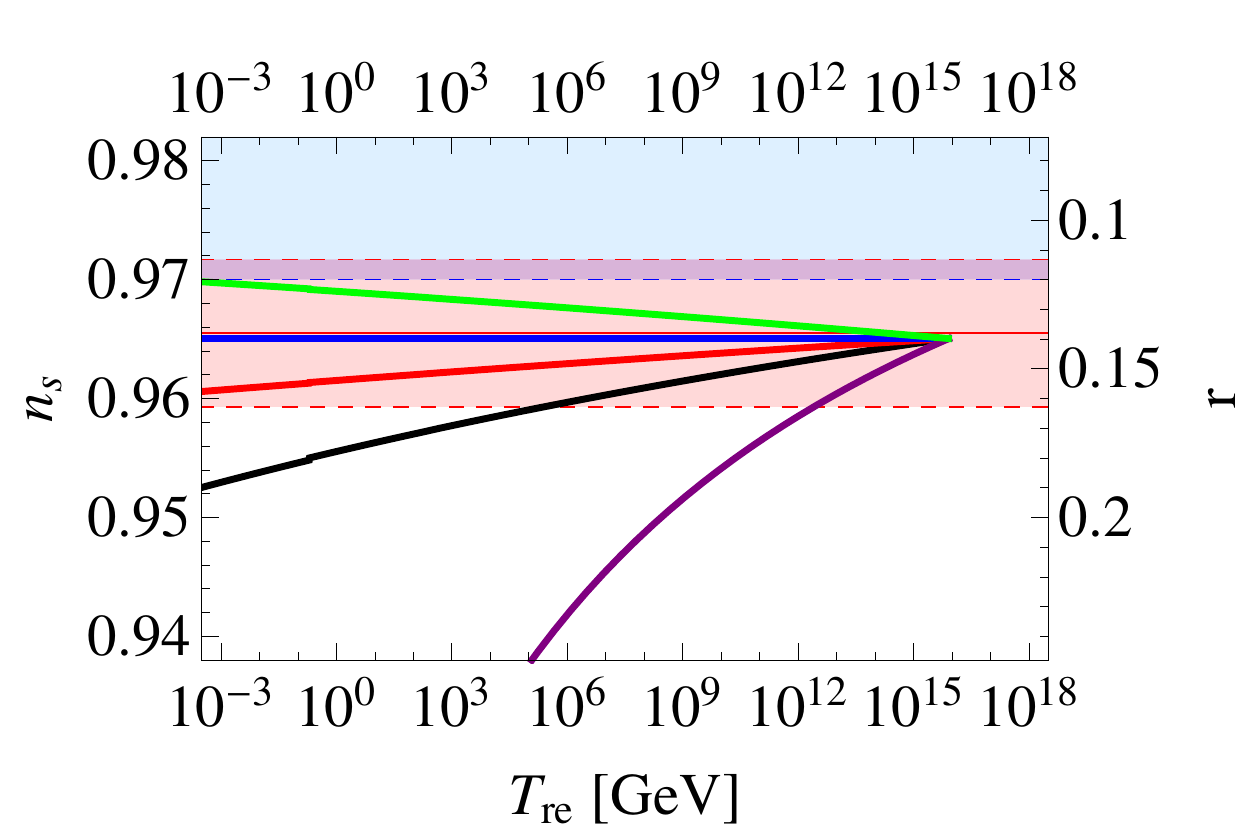}
\caption{The spectral index, $n_s$ (left vertical axis) and the tensor-to-scalar ratio, $r$ (right vertical axis), measured at pivot point $k=0.05\,{\rm Mpc}^{-1}$ as a function of $T_{re}$ for $p=2,$ and $w_{re}=0$ (black), $w_{re}=1/6$ (red), $w_{re}=1/3$ (blue), $w_{re}=-1/3$ (purple), and $w_{re}=2/3$ (green).  To find $N_k,$ we used the same values used in Fig.~\ref{NTPlot}.  The light red area marks the $68\%$ confidence limit range of the Planck 2015 measurement of $n_s$, while the light blue area marks the range of $r$ allowed by the BICEP2/Keck Array and Planck joint analysis.  The light purple area is where these two regions overlap.}
\label{mono2}
\end{center}
\end{figure}

Fig.~\ref{mono2} shows the spectral index, $n_s$ (left vertical axis) and the tensor-to-scalar ratio, $r$ (right vertical axis), measured at pivot point $k=0.05\,{\rm Mpc}^{-1}$ as a function of $T_{re}$ for $p=2,$ and $w_{re}=0$ (black), $w_{re}=1/6$ (red), $w_{re}=1/3$ (blue), $w_{re}=-1/3$ (purple), and $w_{re}=2/3$ (green).  To find $N_k,$ we used the same values used in Fig.~\ref{NTPlot}.  The light red area marks the $68\%$ confidence limit range of the Planck 2015 measurement of $n_s$, while the light blue area marks the range of $r$ allowed by the BICEP2/Keck Array and Planck joint analysis.  The light purple area is where these two regions overlap.

We clarify that we used the central value of $n_s=0.9655$ to calculate the value of the inflaton potential in order to find $N_k$.  $N_k$ is in turn used to calculate $r$ and $n_s$.  However, if $n_s$ is allowed to vary between $0.87$ and $0.97$, as predicted by the models we consider in Fig.~\ref{mono2}, the value of $N_k$ is changed by no more than $0.5$ (i.e. no more than $1.5\%$) for $w_{re}=0$.  

We see that for the well studied $\phi^2$ inflationary potential, while the models reproduce the value of $n_s$ measured by Planck, the upper limit on $r$ found by the BICEP2/Keck Array and Plank joint analysis cannot be satisfied by any of these reheating models.  Even for the exotic $w_{re}=2/3$ model, the predicted value of $r$ does not fall below $r<0.12$ until the predicted reheating temperature drops below the BBN limit of $4\,{\rm MeV}$.  Thus we see that the $\phi ^2$ inflaton potential is no longer favored by current CMB data, in agreement with Fig.~21 of Ref.~\cite{Planck:2015xua}.

 \begin{figure} 
\begin{center}
\includegraphics[width=10cm]{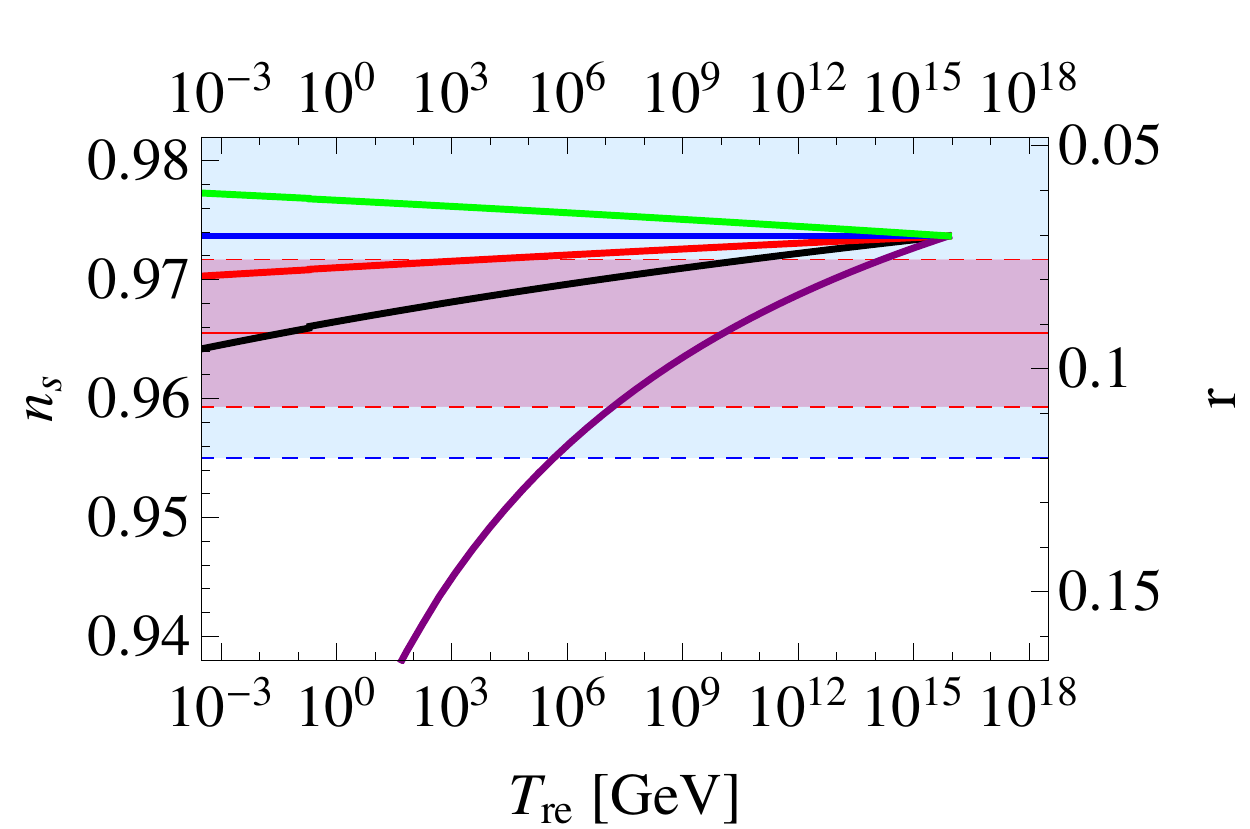}
\caption{Same as Fig.~\ref{mono2}, but for $p=1$.}
\label{mono1}
\end{center}
\end{figure}

\begin{figure} 
\begin{center}
\includegraphics[width=10cm]{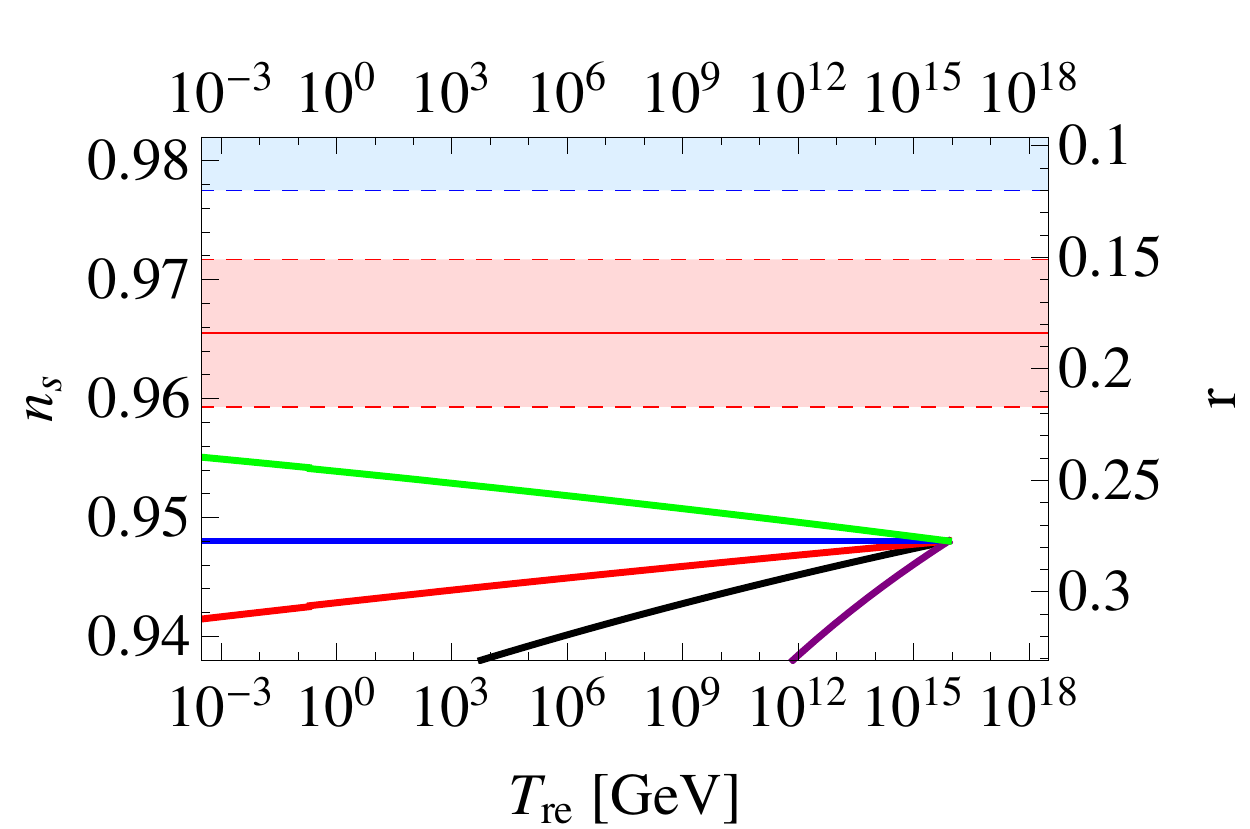}
\caption{Same as Fig.~\ref{mono2}, but for $p=4$.}
\label{mono4}
\end{center}
\end{figure}

Figs.~\ref{mono1} and \ref{mono4} are the same as Fig.~\ref{mono2}, but for $p=1$ and $p=4$, respectively.  We find that a  $\phi^1$ potential fits the CMB measurements better than the $\phi^2$ model.  We see that the $68\%$ confidence limit on $n_s$ implies an upper bound of $T_{re}\lesssim 6\times 10^{10}\,{\rm GeV}$, assuming canonical reheating.  Only exotic reheating models with $w_{re}<0$, such as $w_{re}=-1/3$, allow for higher reheating temperatures.  In fact, with $w_{re}=-1/3$, the $68\%$ confidence limit range in $n_s$ translates to the range $10^7\,{\rm GeV}\lesssim T_{re}\lesssim 2 \times 10^{14}\,{\rm GeV}$ for the reheating temperature.  Instantaneous reheating is not allowed by the CMB measurements.  Models with $w_{re}=1/6$ restrict the reheating temperature to $T_{re}\lesssim 7\times 10^3\,{\rm GeV}$, while models with $w_{re}\geq 1/3$ predict values of $n_s$ that are outside of the Planck 2015 $68\%$ confidence limit range for all possible reheating temperatures.  We find that for a $\phi^4$ potential, the predicted values of $n_s$ and $r$ are outside of the CMB limits for all possible reheating temperatures.

In this section, we have presented bounds on $T_{re}$ for models with power law index $p=1$, $2$, and $4$, and equation of state parameter $w_{re}=0$, $1/6$, $1/3$, $-1/3$, and $2/3$.  However, a similar analysis, making use of Eqs.~\ref{nsma} and \ref{rma}, can be performed for models with other power law indexes and equation of state parameters.

\subsection{Binomial potentials} \label{32}

We have seen in the previous section that instantaneous reheating is not allowed in the $\phi^1$ model.  We now examine analytically, for the first time, the impact of a small departure from this potential by studying binomial inflationary potentials.  By adding a second term to the potential, we wish to find how much of a departure is necessary to allow instantaneous reheating, and then how much of a departure it takes before the lowest possible reheating temperature of $4\,{\rm MeV}$ is excluded.  We thus consider models with inflationary potenials with two terms, taking the form
\begin{equation}
V(\phi)=\lambda M_P^4\left[\left(\frac{\phi}{M_P}\right)^p+b\left(\frac{\phi}{M_P}\right)^q \right].
\label{pot}
\end{equation}
With this potential, the slow roll parameters are 
\begin{equation}
\epsilon_V=\frac{1}{2}\left(\frac{M_P}{\phi}\right)^2\frac{\left[p+bq\left(\frac{\phi}{M_P}\right)^{q-p}\right]^2}{\left[1+b\left(\frac{\phi}{M_P}\right)^{q-p}\right]^2}
\label{epsb}
\end{equation}
and
\begin{equation}
\eta_V=\left(\frac{M_P}{\phi}\right)^2\frac{p(p-1)+bq(q-1)\left(\frac{\phi}{M_P}\right)^{q-p}}{1+b\left(\frac{\phi}{M_P}\right)^{q-p}}.
\label{etab}
\end{equation}
The number of e-folds from when the pivot scale exits the horizon until the end of inflation is
\begin{equation}
N_k=\frac{1}{M_P}\int^{\phi_k}_{\phi_{end}}\frac{\phi}{M_P}\frac{1+b\left(\frac{\phi}{M_P}\right)^{q-p}}{p+bq\left(\frac{\phi}{M_P}\right)^{q-p}}{\rm d}\phi.
\label{Nkb}
\end{equation}

If the first term of the potential dominates over the second term during inflation, i.e.
\begin{equation}
1\gg b\left(\frac{\phi}{M_P}\right)^{q-p}
\end{equation}
and assuming $q/p\sim \mathcal{O}(1)$, Eq.~\ref{Nkb} simplifies to 
\begin{equation}
N_k\simeq\frac{1}{M_P}\int^{\phi_k}_{\phi_{end}}\frac{1}{p}\frac{\phi}{M_P}\left[1+b\left(1-\frac{q}{p}\right)\left(\frac{\phi}{M_P}\right)^{q-p}\right]{\rm d}\phi.
\end{equation}
The solution to this integral is
\begin{equation}
N_k\simeq\frac{1}{2p}\left\{\left(\frac{\phi_k^2-\phi_{end}^2}{M_P^2}\right)+\frac{2b\left(1-\frac{q}{p}\right)}{q-p+2}\left[\left(\frac{\phi_k}{M_P}\right)^{q-p+2}-\left(\frac{\phi_{end}}{M_P}\right)^{q-p+2} \right]\right\}.
\label{Nkapprox}
\end{equation}
The first term on the right side of Eq.~\ref{Nkapprox} is the same as in the case of a monomial potential, and the second term is the first order correction.

As in the case of the monomial potential, we want to find expressions for the slow roll parameters as functions of $N_k$.  If the first term in the potential dominates over the second, Eq.~\ref{epsb} simplifies to
\begin{equation}
\epsilon_V\simeq\frac{p^2}{2}\left(\frac{\phi}{M_P}\right)^{-2}+p^2b\left(\frac{q}{p}-1\right)\left(\frac{\phi}{M_P}\right)^{q-p-2}.
\label{321}
\end{equation}
Rearranging this formula, and again using the approximation that the second term of the potential is subdominant, we find that
\begin{equation}
\left(\frac{\phi}{M_P}\right)^2\simeq \frac{p^2}{2\epsilon_V}+2b\left(\frac{q}{p}-1\right)\left(\frac{p^2}{2\epsilon_V}\right)^{(q-p+2)/2}.
\end{equation}
Using this expression in Eq.~\ref{Nkapprox} and keeping only the first order correction, we find
\begin{equation}
N_k\simeq\frac{1}{2p}\left\{\frac{p^2}{2\epsilon_V}+2b\left(\frac{p^2}{2\epsilon_V}\right)^{(q-p+2)/2}\left(\frac{q}{p}-1\right)\left(1-\frac{1}{q-p+2}\right)\right\}.
\label{eqeps}
\end{equation}

A similar process can be followed for the second slow roll parameter, $\eta_V$
\begin{equation}
N_k\simeq\frac{1}{2p}\left\{\frac{p(p-1)}{\eta_V}+b\left(\frac{p(p-1)}{\eta_V}\right)^{(q-p+2)/2}\left[\frac{q(q-1)-p(p-1)}{p(p-1)}+\frac{2\left(1-\frac{q}{p}\right)}{q-p+2}\right]\right\},
\label{eqeta}
\end{equation}
which is valid for all $p\neq 1$.  In the case of $p=1$, we find instead
\begin{equation}
N_k\simeq\left(\frac{b q(q-1)}{\eta_V}\right)^{2/(3-q)}-\frac{b}{3}\left( \frac{bq(q-1)}{\eta_V}\right)^{3/(3-q)}
\label{eqeta1}
\end{equation}

By inverting Eqs.~\ref{eqeps} and \ref{eqeta} (or \ref{eqeta1} for $p=1$) we can then find an equation for $\epsilon_V$ and $\eta_V$ in terms of the parameter $b$ and $N_k$.  To do this, we must use particular values of $p$ and $q$.  Once we solve these equations, we use Eqs.~\ref{nsm} and \ref{rm} to find the predicted values of $n_s$ and $r$.


\subsection{Study of $V\propto \phi/M_p+b(\phi/M_p)^2$ potentials}

We now focus on the interesting case of a binomial potential with $p=1$ and $q=2$.  We particularly wish to find out how much of a departure from a simple $\phi^1$ monomial potential is necessary to allow instantaneous reheating, and how much of a departure it takes before the lowest possible reheating temperature of $4\,{\rm MeV}$ is excluded by the Planck 2015 $68\%$ confidence limits.

 Before plotting $n_s$ and $r$ as functions of $T_{re}$, it is useful to know which values of $b$ are allowed by the approximation that the second term in the potential is subdominant to the first.  For $p=1$ and $q=2$, we need to satisfy $1\gg b (\phi/M_P)^1$ during inflation.  Keeping the leading term in Eq.~\ref{321}, $(\phi/M_P)^2\simeq p^2/(2\epsilon_V)$ and using Eq.~\ref{eqeps}, even with $N_k$ at its maximum value of $57$ (assuming canonical reheating), we find that $b\lesssim10^{-1}$ is enough to ensure that the second term in the potential is subdominant to the first.

 \begin{figure} 
\begin{center}
\includegraphics[width=7.5cm]{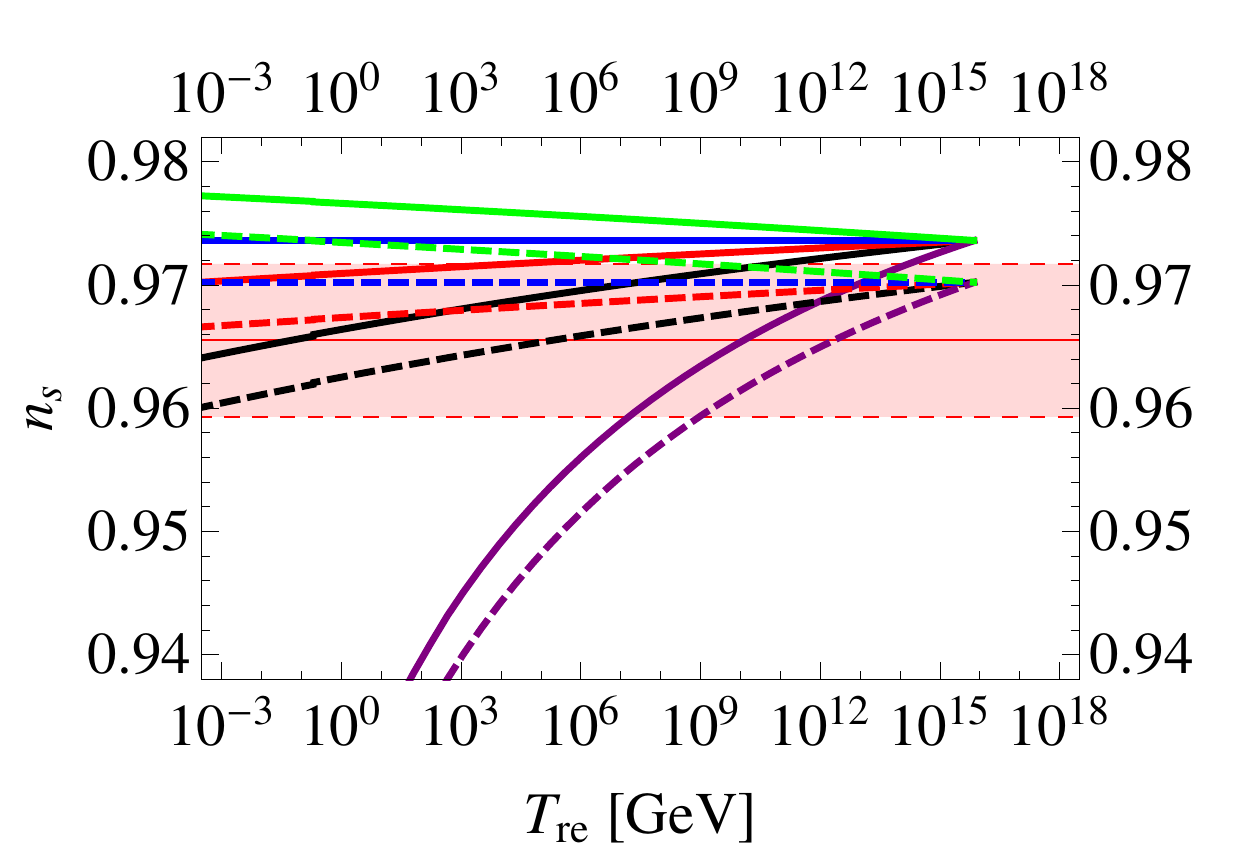}$~~~$
\includegraphics[width=7.5cm]{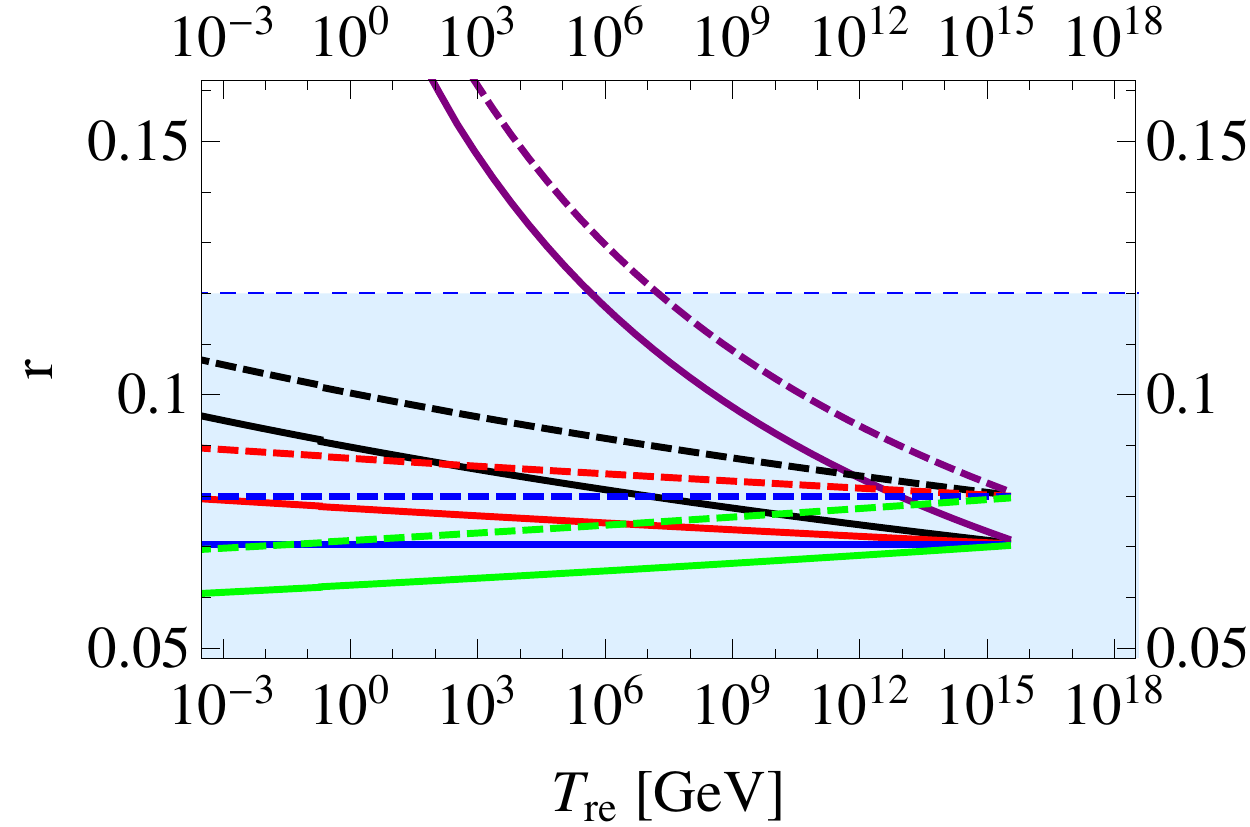}
\caption{The spectral index, $n_s$ (left) and the tensor-to-scalar ratio, $r$ (right), measured at pivot scale $k=0.05\,{\rm Mpc}^{-1}$ as a function of $T_{re}$ for a binomial potential with $p=1$, $q=2$, $b=10^{-2}$, and $w_{re}=0$ (black dashed), $w_{re}=1/6$ (red dashed), $w_{re}=1/3$ (blue dashed), $w_{re}=-1/3$ (purple dashed), and $w_{re}=2/3$ (green dashed).  The solid black, red, blue, purple, and green lines show $n_s$ and $r$ for the corresponding monomial potential with $p=1$.  To find $N_k,$ we used the same values used in Fig.~\ref{NTPlot}.  The red and blue shaded areas mark the $68\%$ confidence limit region allowed by the Planck 2015 measurement of the spectral index and the BICEP2/Keck Array and Planck joint analysis upper bound on the tensor-to-scalar ratio, respectively.  We are no longer able to plot both $n_s$ and $r$ on the same plot, since the relationship between the two now depends on the parameter $b$.}
\label{bi}
\end{center}
\end{figure}

Fig.~\ref{bi} shows the spectral index, $n_s$ (left) and the tensor-to-scalar ratio, $r$ (right), measured at pivot scale $k=0.05\,{\rm Mpc}^{-1}$ as a function of $T_{re}$ for a binomial potential with $p=1$, $q=2$, $b=10^{-2}$, and $w_{re}=0$ (black dashed), $w_{re}=1/6$ (red dashed), $w_{re}=1/3$ (blue dashed), $w_{re}=-1/3$ (purple dashed), and $w_{re}=2/3$ (green dashed).  The solid black, red, blue, purple, and green lines show $n_s$ and $r$ for the corresponding monomial potential with $p=1$.  To find $N_k,$ we used the same values used in Fig.~\ref{NTPlot}.  The red and blue shaded areas mark the $68\%$ confidence limit region allowed by the Planck 2015 measurement of the spectral index and the BICEP2/Keck Array and Planck joint analysis upper bound on the tensor-to-scalar ratio, respectively.  We are no longer able to plot both $n_s$ and $r$ on the same plot, since the relationship between the two now depends on the parameter $b$.  As expected, when the second term in the potential is subdominant to the first, the dashed lines closely follow the solid lines for a monomial potential with $p=1$, but are slightly offset in the direction of the lines for a monomial potential with $p=2$.  As $b$ increases, the dashed lines move increasingly farther from the solid lines, and toward the lines for a monomial potential with $p=2$.   

In particular, we want to understand how the limits on $T_{re}$ change as $b$ increases.  The range of values of $T_{re}$ with predicted values of $r$ and $n_s$ that fall within the ranges given in Eqs.~\ref{rval} and \ref{nsval}, respectively, can be read off of Fig.~\ref{bi}.  In the case of canonical reheating ($w_{re}=0$), with $p=1$, $q=2$, and $b=10^{-2}$, the constraints are  relaxed, compared to those of a monomial potential with $p=1$, because the addition of the second term in the potential lowers the predicted value of $n_s$.  In this case, we see that instantaneous reheating is allowed (as compared with an upper limit of  $T_{re}<6\times 10^{10}\,{\rm GeV}$ for a pure monomial potential with $p=1$).  The lowest reheating temperature of $4\,{\rm Mev}$ is also allowed in this case.

Finally, we can find how much deviation from a monomial potential with $p=1$ is necessary to allow instantaneous reheating, and how much of a depature it takes before the lowest possible reheating temperature is no longer allowed.  We find that in the case of binomial potentials with $p=1$ and $q=2$, as the subdominant term increases (i.e. $b$ gets larger), instantaneous reheating is allowed for the first time when $b=5.3\times 10^{-3}$, when the predicted value of $n_s$ falls within the $68\%$ confidence limit range of the Planck 2015 measurement.  Assuming canonical reheating, the lowest reheating temperature of $4\,{\rm MeV}$ is also allowed by this value of $b$.  As the subdominant term continues to increase, the predicted value of $n_s$ at $T_{re}=4\,{\rm MeV}$ first falls out of the $68\%$ confidence limit range of the Planck measurement when $b=1.4\times 10^{-2}$.  This means that when $b$ takes values between $5.3\times 10^{-3}<b<1.4\times 10^{-2}$, all possible values of the reheating temperature are allowed by the Planck 2015 $68\%$ confidence limit range on $n_s$ and the BICEP2/Keck Array and Planck joint analysis upper bound on $r$, assuming canonical reheating.

While we have considered the interesting case of a binomial potential with $p=1$ and $q=2$, a similar analysis for inflaton potentials with different values of $p$ and $q$ can be carried out using the procedure described in Section~\ref{32}.  Corresponding limits on the reheating temperature of the Universe can then be found for these models as well.

\section{Conclusions}

In this paper we rederived the relationship between the number of e-folds from when the pivot scale exits the horizon until the end of inflation and the reheating temperature.  We find that in the case of canonical reheating, and a quadratic inflationary potential, the lower bound on $N_k$, measured at pivot scale $k=0.05\,{\rm Mpc}^{-1}$, is $N_k\gtrsim 42.5$.  In addition, for reheating scenarios with $w_{re}\leq 1/3$, $N_k$ cannot exceed $57$.  If  our assumption that the energy density of the Universe remains constant during inflation does not hold, then the upper limit will be raised.  For example, if the energy density falls by a factor of $\rho_k/\rho_{end}\simeq 3700$ between when the pivot scale left the horizon and the end of inflation, as in the most quickly falling potential we considered, then the upper limit will be raised by $\Delta N_k\simeq 2.7$ e-folds, for $w_{re}=0$.

We then found the dependence of the spectral index, $n_s$, and the tensor-to-scalar ratio, $r$, on $N_k$ (and thereby $T_{re}$) for the archetypal models of inflation, large field inflation models, which are characterized by a monomial potential.  We also studied, as a modification of the monomial potential, binomial inflationary potentials in which one term dominates over the other.  We used the measurement of $n_s$ made by Planck 2015 (given in Eq.~\ref{nsval}) and the upper limit on $r$ given by the joint analysis of the BICEP2/Keck Array and Planck experiments (given in Eq.~\ref{rval}) to constrain the possible reheating temperatures of these models.  We see that for the well studied $\phi^2$ inflationary potential, while the models reproduce the value of $n_s$ favored by Planck, the upper limit on $r$ found by the BICEP2/Keck Array and Plank joint analysis cannot be satisfied by any of the reheating models we consider, in agreement with Fig.~21 of Ref~\cite{Planck:2015xua}.  Even for the exotic $w_{re}=2/3$ model, the predicted value of $r$ does not fall below $r<0.12$ until the predicted reheating temperature drops below the BBN limit of $4\,{\rm MeV}$.  Thus we find that the $\phi ^2$ inflaton potential is no longer favored by current CMB data.

 We see that a  $\phi^1$ potential provides a good fit to the CMB measurements.  We find that the Planck 2015 $68\%$ confidence limit range on $n_s$ implies an upper bound of $T_{re}\lesssim 6\times 10^{10}\,{\rm GeV}$, assuming canonical reheating.  Only exotic reheating models with $w_{re}<0$, such as $w_{re}=-1/3$, allow for higher reheating temperatures.  In fact, with $w_{re}=-1/3$, the $68\%$ confidence limit range in $n_s$ translates to the range $10^7\,{\rm GeV}\lesssim T_{re}\lesssim 2 \times 10^{14}\,{\rm GeV}$ for the reheating temperature.  Instantaneous reheating is not allowed by the CMB measurements.  Models with $w_{re}=1/6$ restrict the reheating temperature to $T_{re}\lesssim 7\times 10^3\,{\rm GeV}$, while models with $w_{re}\geq 1/3$ predict values of $n_s$ that are outside of the $68\%$ confidence limit range of the Planck measurement for all possible reheating temperatures.  We see that for a $\phi^4$ potential, the predicted values of $n_s$ and $r$ are outside of the CMB limits for all possible reheating temperatures.

We then explored, for the first time, inflationary models with binomial potentials ($V\propto \phi^p+b\phi^q$) for general $p$ and $q$, and use our results to find how much deviation from a monomial potential with $p=1$ is necessary to allow instantaneous reheating, and how much of a depature it takes before the lowest possible reheating temperature is no longer allowed.  We find that in the case of binomial potentials with $p=1$ and $q=2$, as the subdominant ($\phi^2$) term increases (i.e. $b$ gets larger), instantaneous reheating is allowed for the first time when $b=5.3\times 10^{-3}$, when the predicted value of $n_s$ falls within the $68\%$ confidence limit range of the Planck 2015 measurement.  Assuming canonical reheating, the lowest reheating temperature of $4\,{\rm MeV}$ is also allowed by this value of $b$.  As the subdominant term continues to increase, the predicted value of $n_s$ at $T_{re}=4\,{\rm MeV}$ first falls out of the $68\%$ confidence limit range of the Planck measurement when $b=1.4\times 10^{-2}$.  This means that when $b$ takes values between $5.3\times 10^{-3}<b<1.4\times 10^{-2}$, all possible values of the reheating temperature are allowed by the Planck 2015 $68\%$ confidence limit range on $n_s$ and the BICEP2/Keck Array and Planck upper bound on $r$, assuming canonical reheating.

We find particularly interesting that for a monomial potential with $p=1$, the $68\%$ confidence limits on $n_s$ and the upper bound on $r$ imply an upper bound $T_{re}\lesssim 6\times 10^{10}$ for the reheating temperature, assuming canonical reheating.  If the Universe has a low reheating temperature, like those allowed in this model, the possibility is opened that dark matter could be produced during the reheating period instead of when the Universe is radiation dominated, which could lead to very different predictions for the relic density and even momentum distribution of WIMPs, sterile neutrinos, and axions from the predictions assuming that dark matter is produced when the Universe is radiation dominated.

\section*{Acknowledgements}

T.R. and G.G. were supported in part by the  Department of Energy under Award Number DE-SC0009937.  This research was also supported in part by the National Science Foundation under Grant No. PHY11-25915 (through the Kavli Institute for Theoretical Physics, KITP, at the University of California, Santa Barbara, where G.G. carried out part of the work).


\begin{thebibliography}{99}

\bibitem{Hannestad:2004px} 
  S.~Hannestad,
  {\it What is the lowest possible reheating temperature?,}
 {\it Phys.\ Rev.\ D} {\bf 70}, 043506 (2004)
  [astro-ph/0403291].

\bibitem{DeBernardis:2008zz} 
  F.~De Bernardis, L.~Pagano and A.~Melchiorri,
  {\it New constraints on the reheating temperature of the universe after WMAP-5,}
{\it  Astropart.\ Phys.\ } {\bf 30}, 192 (2008).
  M.~Kawasaki, K.~Kohri and N.~Sugiyama,
  {\it MeV scale reheating temperature and thermalization of neutrino background,}
 {\it Phys.\ Rev.\ D} {\bf 62}, 023506 (2000)
  [astro-ph/0002127].
  M.~Kawasaki, K.~Kohri and N.~Sugiyama,
  {\it Cosmological constraints on late time entropy production,}
 {\it Phys.\ Rev.\ Lett.\ } {\bf 82}, 4168 (1999)
  [astro-ph/9811437].

\bibitem{Gelmini:2006pw}
 G.Giudice, E.Kolb, and A.Riotto
 {\it Largest temperature of the radiation era and its cosmological implications},
 {\it Phys.\ Rev.\  D} {\bf 64} (2001)  023508
  [arXiv:hep-ph/0005123];
  N.Fornengo, A.Riotto, and S.Scopel
 {\it Supersymmetric Dark Matter and the Reheating Temperature of the Universe},
 {\it Phys.\ Rev.\  D} {\bf 67} (2003)  023514
  [arXiv:hep-ph/0208072];
  C.Pallis
 {\it Massive Particle Dacay and Cold Dark Matter Abundance},
 {\it Astorpart. Phys.} {\bf 21} (2004)  689
  [arXiv:hep-ph/0402033];
  G.Gelmini and P.Gondolo,
 {\it Neutralino with the right cold dark matter abundance in (almost) any
  supersymmetric model},
 {\it Phys.\ Rev.\  D} {\bf 74} (2006)  023510
 [arXiv:hep-ph/0602230];
  M.~Drees, H.~Iminniyaz and M.~Kakizaki,
  {\it Abundance of cosmological relics in low-temperature scenarios,}
 {\it Phys.\ Rev.\  D} {\bf 73} (2006) 123502 [arXiv:hep-ph/0603165];
G. Gelmini {\it et al}
  {\it The effect of a late decaying scalar on the neutralino relic density,}
 {\it Phys.  Rev. D} {\bf 74} (2006)083514 
  [hep-ph/0605016].
  G.~B.~Gelmini, J.~H.~Huh and T.~Rehagen,
  {\it ``Asymmetric dark matter annihilation as a test of non-standard cosmologies,''
  JCAP }{\bf 1308}, 003 (2013)
  [arXiv:1304.3679 [hep-ph]].
  L.~Visinelli and P.~Gondolo,
  {\it ``Axion cold dark matter in non-standard cosmologies,''
  Phys.\ Rev.\ D} {\bf 81}, 063508 (2010)
  [arXiv:0912.0015 [astro-ph.CO]].
  G.~Gelmini, S.~Palomares-Ruiz and S.~Pascoli,
  {\it ``Low reheating temperature and the visible sterile neutrino,''
  Phys.\ Rev.\ Lett.\ } {\bf 93}, 081302 (2004)
  [astro-ph/0403323].
  G.~Gelmini, E.~Osoba, S.~Palomares-Ruiz and S.~Pascoli,
  {\it ``MeV sterile neutrinos in low reheating temperature cosmological scenarios,''
  JCAP }{\bf 0810}, 029 (2008)
  [arXiv:0803.2735 [astro-ph]].
  C.~E.~Yaguna,
  {\it ``Sterile neutrino production in models with low reheating temperatures,''
  JHEP }{\bf 0706}, 002 (2007)
  [arXiv:0706.0178 [hep-ph]].
  T.~Rehagen and G.~B.~Gelmini,
  {\it ``Effects of kination and scalar-tensor cosmologies on sterile neutrinos,''
  JCAP }{\bf 1406}, 044 (2014)
  [arXiv:1402.0607 [hep-ph]].
  L.~Roszkowski, S.~Trojanowski and K.~Turzyński,
  {\it``Neutralino and gravitino dark matter with low reheating temperature,''
  JHEP} {\bf 1411}, 146 (2014)
  [arXiv:1406.0012 [hep-ph]].


\bibitem{Gelmini:2006vn} 
  G.~Gelmini and C.~E.~Yaguna
  {\it Constraints on Minimal SUSY models with warm dark matter neutralinos,}
  {\it Phys.\ Lett.\ B} {\bf 643} (2006) 241
  [hep-ph/0607012];
  G.~B.~Gelmini and P.~Gondolo
  {\it Ultra-cold WIMPs: relics of non-standard pre-BBN cosmologies,}
  {\it JCAP} {\bf 0810} (2008) 002
  [astro-ph/0803.2349].  

\bibitem{Ade:2014xna} 
  P.~A.~R.~Ade {\it et al.}  [BICEP2 Collaboration],
  {\it ``Detection of B-Mode Polarization at Degree Angular Scales by BICEP2,''
  Phys.\ Rev.\ Lett.\ } {\bf 112}, 241101 (2014)
  [arXiv:1403.3985 [astro-ph.CO]].

\bibitem{Ade:2015fwj} 
  P.~A.~R.~Ade {\it et al.}  [BICEP2 and Keck Array Collaborations],
 {\it ``BICEP2 / Keck Array V: Measurements of B-mode Polarization at Degree Angular Scales and 150 GHz by the Keck Array,''}
  arXiv:1502.00643 [astro-ph.CO].

\bibitem{Ade:2015tva} 
  P.~A.~R.~Ade {\it et al.}  [BICEP2 and Planck Collaborations],
  {\it ``A Joint Analysis of BICEP2/Keck Array and Planck Data,''
  Submitted to: Phys.Rev.Lett.}
  [arXiv:1502.00612 [astro-ph.CO]].

\bibitem{Guth:1980zm} 
  A.~H.~Guth,
{\it  ``The Inflationary Universe: A Possible Solution to the Horizon and Flatness Problems,''}
 {\it Phys.\ Rev.\ D} {\bf 23}, 347 (1981).
  A.~D.~Linde,
  {\it``A New Inflationary Universe Scenario: A Possible Solution of the Horizon, Flatness, Homogeneity, Isotropy and Primordial Monopole Problems,''}
 {\it Phys.\ Lett.\ B} {\bf 108}, 389 (1982).
  A.~Albrecht and P.~J.~Steinhardt,
  {\it``Cosmology for Grand Unified Theories with Radiatively Induced Symmetry Breaking,''}
{\it  Phys.\ Rev.\ Lett.}  {\bf 48}, 1220 (1982).

\bibitem{Linde:1983gd} 
  A.~D.~Linde,
{\it  ``Chaotic Inflation,''}
 {\it Phys.\ Lett.\ B} {\bf 129}, 177 (1983).

\bibitem{Allahverdi:2010xz} 
  R.~Allahverdi, R.~Brandenberger, F.~Y.~Cyr-Racine and A.~Mazumdar,
  {\it``Reheating in Inflationary Cosmology: Theory and Applications,''
  Ann.\ Rev.\ Nucl.\ Part.\ Sci.\ } {\bf 60}, 27 (2010)
  [arXiv:1001.2600 [hep-th]].

\bibitem{Martin:2006rs} 
  J.~Martin and C.~Ringeval,
  JCAP {\bf 0608}, 009 (2006)
  [astro-ph/0605367].

\bibitem{Martin:2010kz} 
  J.~Martin and C.~Ringeval,
  {\it``First CMB Constraints on the Inflationary Reheating Temperature,''
  Phys.\ Rev.\ D} {\bf 82}, 023511 (2010)
  [arXiv:1004.5525 [astro-ph.CO]].

\bibitem{Martin:2014vha} 
  J.~Martin, C.~Ringeval and V.~Vennin,
  {\it ``Encyclopædia Inflationaris,''
  Phys.\ Dark Univ.\ } (2014)
  [arXiv:1303.3787 [astro-ph.CO]].

\bibitem{Dai:2014jja} 
  L.~Dai, M.~Kamionkowski and J.~Wang,
{\it  ``Reheating constraints to inflationary models,''}
  Phys.\ Rev.\ Lett.\  {\bf 113}, 041302 (2014)
  [arXiv:1404.6704 [astro-ph.CO]].

\bibitem{Drewes:2013iaa} 
  M.~Drewes and J.~U.~Kang,
  {\it ``The Kinematics of Cosmic Reheating,''
  Nucl.\ Phys.\ B} {\bf 875}, 315 (2013)
  [Nucl.\ Phys.\ B {\bf 888}, 284 (2014)]
  [arXiv:1305.0267 [hep-ph]].
  M.~Drewes,
  {\it ``On finite density effects on cosmic reheating and moduli decay and implications for Dark Matter production,''
  JCAP }{\bf 1411}, no. 11, 020 (2014)
  [arXiv:1406.6243 [hep-ph]].

\bibitem{Belinsky:1985zd} 
  V.~A.~Belinsky, I.~M.~Khalatnikov, L.~P.~Grishchuk and Y.~B.~Zeldovich,
  {\it``Inflationary Stages In Cosmological Models With A Scalar Field,''
  Phys.\ Lett.\ B} {\bf 155}, 232 (1985).

\bibitem{Creminelli:2014oaa} 
  P.~Creminelli, D.~López Nacir, M.~Simonović, G.~Trevisan and M.~Zaldarriaga,
 {\it ``$\phi^2$ or Not $\phi^2$: Testing the Simplest Inflationary Potential,''
  Phys.\ Rev.\ Lett.\  }{\bf 112}, 241303 (2014)
  [arXiv:1404.1065 [astro-ph.CO]].
  P.~Creminelli, D.~L.~Nacir, M.~Simonović, G.~Trevisan and M.~Zaldarriaga,
 {\it ``$\phi^2$ Inflation at its Endpoint,''}
  arXiv:1405.6264 [astro-ph.CO].

\bibitem{McAllister:2008hb} 
  L.~McAllister, E.~Silverstein and A.~Westphal,
  {\it``Gravity Waves and Linear Inflation from Axion Monodromy,''
  Phys.\ Rev.\ D} {\bf 82}, 046003 (2010)
  [arXiv:0808.0706 [hep-th]].

\bibitem{Turner:1983he}
  M.~S.~Turner,
  {\it``Coherent Scalar Field Oscillations in an Expanding Universe,''
  Phys.\ Rev.\ D }{\bf 28} (1983) 1243.

\bibitem{Abbott:1982hn} 
  L.~F.~Abbott, E.~Farhi and M.~B.~Wise,
  {\it ``Particle Production in the New Inflationary Cosmology,''
  Phys.\ Lett.\ B }{\bf 117}, 29 (1982).

\bibitem{Dolgov:1982th} 
  A.~D.~Dolgov and A.~D.~Linde,
  {\it``Baryon Asymmetry in Inflationary Universe,''
  Phys.\ Lett.\ B} {\bf 116}, 329 (1982).

\bibitem{Albrecht:1982mp} 
  A.~Albrecht, P.~J.~Steinhardt, M.~S.~Turner and F.~Wilczek,
  {\it``Reheating an Inflationary Universe,''
  Phys.\ Rev.\ Lett.\ } {\bf 48}, 1437 (1982).

\bibitem{Podolsky:2005bw} 
  D.~I.~Podolsky, G.~N.~Felder, L.~Kofman and M.~Peloso,
  {\it ``Equation of state and beginning of thermalization after preheating,''
  Phys.\ Rev.\ D} {\bf 73}, 023501 (2006)
  [hep-ph/0507096].
  J.~F.~Dufaux, G.~N.~Felder, L.~Kofman, M.~Peloso and D.~Podolsky,
  {\it``Preheating with trilinear interactions: Tachyonic resonance,''
  JCAP }{\bf 0607}, 006 (2006)
  [hep-ph/0602144].

\bibitem{Mukhanov:1981xt} 
  V.~F.~Mukhanov and G.~V.~Chibisov,
  {\it ``Quantum Fluctuation and Nonsingular Universe. (In Russian),''
  JETP Lett.\ } {\bf 33}, 532 (1981)
  [Pisma Zh.\ Eksp.\ Teor.\ Fiz.\  {\bf 33}, 549 (1981)].
  S.~W.~Hawking,
  {\it``The Development of Irregularities in a Single Bubble Inflationary Universe,''
  Phys.\ Lett.\ B} {\bf 115}, 295 (1982).
  A.~A.~Starobinsky,
  {\it``Dynamics of Phase Transition in the New Inflationary Universe Scenario and Generation of Perturbations,''
  Phys.\ Lett.\ B} {\bf 117}, 175 (1982).
  A.~H.~Guth and S.~Y.~Pi,
  {\it``Fluctuations in the New Inflationary Universe,''
  Phys.\ Rev.\ Lett.\ } {\bf 49}, 1110 (1982).
  J.~M.~Bardeen, P.~J.~Steinhardt and M.~S.~Turner,
  {\it``Spontaneous Creation of Almost Scale - Free Density Perturbations in an Inflationary Universe,''
  Phys.\ Rev.\ D} {\bf 28}, 679 (1983).

\bibitem{Planck:2015xua} 
  P.~A.~R.~Ade {\it et al.}  [Planck Collaboration],
  {\it ``Planck 2015 results. XIII. Cosmological parameters,''}
  arXiv:1502.01589 [astro-ph.CO].

\bibitem{Munoz:2014eqa} 
  J.~B.~Munoz and M.~Kamionkowski,
 {\it ``Equation-of-State Parameter for Reheating,''
  Phys.\ Rev.\ D }{\bf 91}, no. 4, 043521 (2015)
  [arXiv:1412.0656 [astro-ph.CO]].
  J.~Martin, C.~Ringeval and V.~Vennin,
  {\it ``Observing Inflationary Reheating,''
  Phys.\ Rev.\ Lett.\  }{\bf 114}, no. 8, 081303 (2015)
  [arXiv:1410.7958 [astro-ph.CO]].
  J.~Martin,
 {\it ``The Observational Status of Cosmic Inflation after Planck,''}
  arXiv:1502.05733 [astro-ph.CO].
  J.~L.~Cook, E.~Dimastrogiovanni, D.~A.~Easson and L.~M.~Krauss,
 {\it ``Reheating predictions in single field inflation,''}
  arXiv:1502.04673 [astro-ph.CO].

\bibitem{Domcke:2015iaa} 
  V.~Domcke and J.~Heisig,
  {\it ``Constraints on the reheating temperature from sizable tensor modes,''}
  arXiv:1504.00345 [astro-ph.CO].

\bibitem{Beringer}
J.~Beringer et al. (Particle Data Group),
{\it Phys. Rev. D}{\bf 86}, 010001 (2012)


\end{thebibliography}
\end{document}